\documentclass{PoS}

\usepackage{amsfonts, amsmath, amsthm, amssymb} 
\usepackage{multirow}
\usepackage{hhline}

\usepackage[english]{babel}
\usepackage{cite}
\usepackage{ifpdf}
\usepackage{subfigure}

\usepackage{graphicx} 
\graphicspath{{figures/}} 

\usepackage{booktabs} 
\usepackage[font=small,labelfont=bf]{caption} 
\usepackage{epsf}
\usepackage{rotating}
\usepackage{graphicx}
\usepackage{amsmath}
\usepackage{fancyhdr}
\usepackage{lineno}
\usepackage{dutchcal}
\usepackage{babel}
\usepackage{graphics}
\usepackage{pstricks}
\usepackage{color}
\usepackage{multirow}

\newcommand{\lsim}{\mathrel{\mathop{\kern 0pt \rlap
  {\raise.2ex\hbox{$<$}}}
  \lower.9ex\hbox{\kern-.190em $\sim$}}}
\newcommand{\gsim}{\mathrel{\mathop{\kern 0pt \rlap
  {\raise.2ex\hbox{$>$}}}
  \lower.9ex\hbox{\kern-.190em $\sim$}}}

\newcommand{\be}{\begin{equation}}
\newcommand{\ee}{\end{equation}}
\newcommand{\bea}{\begin{eqnarray}}
\newcommand{\eea}{\end{eqnarray}}

\def\ptmiss{\not\!\!{p_T}}

\def\ptmiss{\not\!\!{p_T}}

\title{Probing and Distinguishing Representations at the LHC}

\ShortTitle{}

\author{\speaker{Antonio Costantini}\\
        University of Salento - INFN Lecce\\
        E-mail: \email{antonio.costantini@le.infn.it}}

\abstract{The discovery made at the Large Hadron Collider (LHC) has revealed that the spontaneous symmetry breaking mechanism is realised in a gauge theory such as the Standard Model (SM) by at least one Higgs doublet. However, the possible existence of other scalar bosons cannot be excluded. We analyze signatures extensions of the SM, characterized by an extra representations of scalars, in view of the recent and previous Higgs data. We show that such representations can be probed and distinguished, mostly with multileptonic final states, with a relatively low luminosity at the LHC.}

\FullConference{Sixth Annual Conference on Large Hadron Collider Physics (LHCP2018)\\
                4-9 June 2018\\
                Bologna, Italy}

\begin{document}

\section{Triplet-Singlet Extended MSSM}

In 2012 the ATLAS \cite{ATLASdisc} and CMS \cite{CMSdisc} collaborations announced the discovery of a 125 GeV mass resonance with the properties that are mostly of the SM Higgs boson. The discovery made at the Large Hadron Collider (LHC) has revealed that the spontaneous symmetry breaking (SSB) mechanism is realised in a gauge theory such as the SM by at least one Higgs doublet. However, the possible existence of other scalar bosons cannot be excluded. 

We consider an extension of the MSSM with a (gauge) singlet superfield $\hat S$ and a triplet superfield $\hat T$ with $Y=0$
\begin{equation}\label{spf}
 \hat T = \begin{pmatrix}
       \sqrt{\frac{1}{2}}\hat T^0 & \hat T_2^+ \cr
      \hat T_1^- & -\sqrt{\frac{1}{2}}\hat T^0
       \end{pmatrix},\qquad \hat{H}_u= \begin{pmatrix}
      \hat H_u^+  \cr
       \hat H^0_u
       \end{pmatrix},\qquad \hat{H}_d= \begin{pmatrix}
      \hat H_d^0  \cr
       \hat H^-_d
       \end{pmatrix},\qquad \hat S.
 \end{equation}
dubbed the TNMSSM \cite{BC1, BC2, BC3, BC4}. The superpotetial is cubic, similarly to the NMSSM case.
The feature of having a cubic superpotential play a crucial role in the phenomenology of both neutral and charged extra states, as detailed in \cite{BC2, BC3, BC4}.

\subsection{Pseudoscalar Higgs}

\begin{figure}
\begin{center}
\mbox{\subfigure[]{\includegraphics[width=0.16\linewidth]{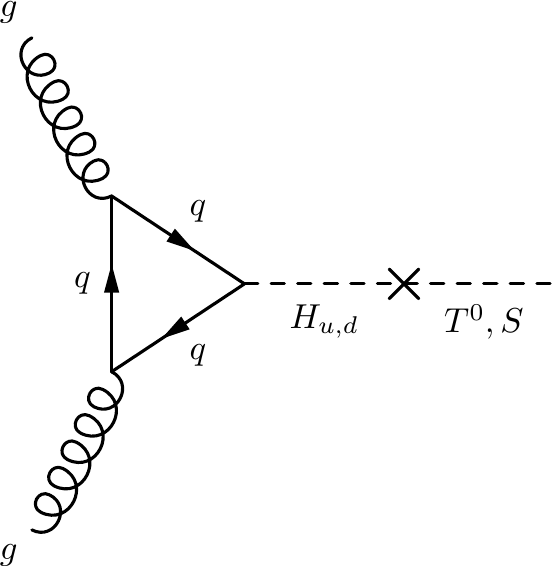}}\hspace{1cm}
\subfigure[]{\includegraphics[width=0.18\linewidth]{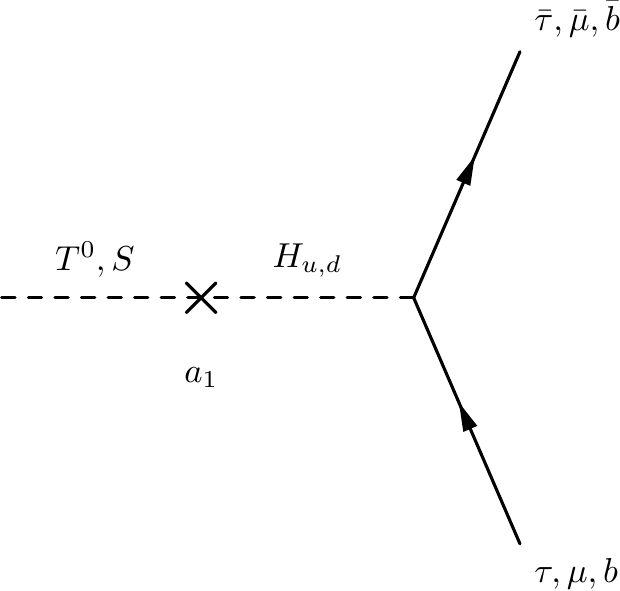}}\hspace{2cm}
\subfigure[]{\includegraphics[width=0.2\linewidth]{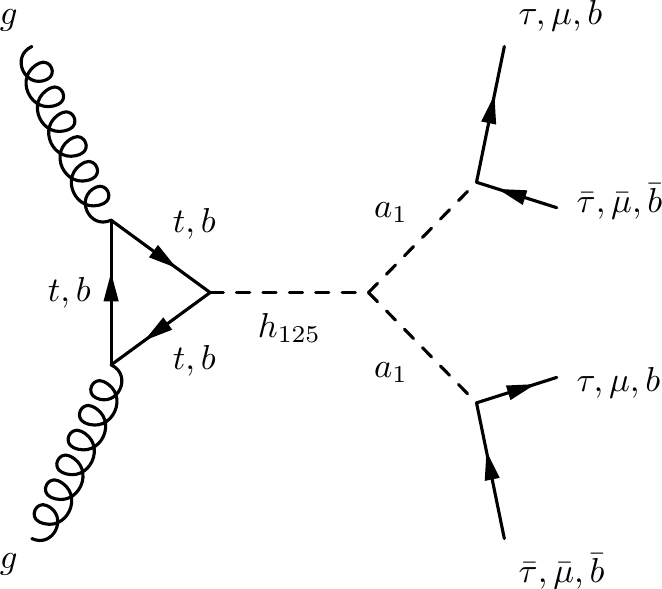}}}
\captionof{figure}{Feynman diagrams for the pair production of a singlet-like light pseudoscalar at the LHC.}\label{pseudo}
\end{center}
\end{figure}
The discovered Higgs boson $h_{125}$ can decay into two light pseudoscalars, as shown in Figure \ref{pseudo}, which further decay into $\tau$ or $b$ pairs. We select various final states in order to prove the existence of the light pseudoscalar \cite{BC2}.
We ask for a final state with $n_j\leq 5$, in which we demand the presence of at least two $b_{\rm{jet}}$'s and two $\tau_{\rm{jet}}$'s and we have also required that the missing transverse momentum is smaller than 30 GeV 
$(\&\,\ptmiss \leq 30 \, \rm{GeV})$.
In addition we apply some other cuts on the signal ($|m_{bb}-m_Z|>10$ GeV, $|m_{bb}-m_{h_{125}}|>10$ GeV, $|m_{\tau\tau}-m_Z|>10$ GeV, $|m_{\tau\tau}-m_{h_{125}}|>10$ GeV, $m_{\tau\tau}<125$ GeV and $m_{bb}<125$ GeV). The most dominant SM backgrounds are those from $t\bar{t}$, $ZZ$, $Zh$, $b\bar{b}h$ and $b\bar{b}Z$ respectively. The signal significances are  $4.47 \,\sigma$, $10.18\, \sigma$ and $5.98 \,\sigma$ respectively for BP1, BP2 and BP3.
Next we have analyzed the invariant mass distributions of the $b_{\rm{jet}}$ pair selecting events with $|m_{bb}-m_{a_1}|\leq 10$ GeV. At an integrated luminosity of 100 fb$^{-1}$, and at a center of mass energy of 14 TeV, the significances are now $8.70\, \sigma$, $7.74 \,\sigma$ and $9.79 \,\sigma$ in the three cases.
For the $\tau_{\rm{jet}}$ case we found, at a centre of mass energy of 14 TeV $14.78\, \sigma$, $6.56 \,\sigma$ and $12.93 \,\sigma$. 
Another signal considered is the final state characterized by $ n_j\leq 5\,[\geq 3\tau_{\rm{jet}}].$
We then add some further kinematical cuts ($|m_{\tau\tau}-m_Z|>10$ GeV, $m_{\tau\tau}\leq125$ GeV, $p_T^{\tau_{j_1}}\leq 100\, \&\, p_T^{\tau_{j_{2,3}}}\leq 50$ GeV).
Without the mass peak analysis the signal significance at an integrated luminosity of 100 fb$^{-1}$ and at center of mass energy of 14 TeV are $3.79 \,\sigma$, $8.38 \, \sigma$ and $5.81\, \sigma$.
Around the pseudoscalar mass peak ($p_1:|m_{\tau\tau}-m_{a_1}|\leq 10$ GeV) with an integrated luminosity of 100 fb$^{-1}$ and at 14 TeV e.c.m. the signal significance are $5.16\, \sigma$, $4.00 \,\sigma$ and $6.03\, \sigma$.

\subsection{Charged Higgs}

\begin{figure}
\begin{center}
\mbox{\subfigure[]{\includegraphics[width=0.15\linewidth]{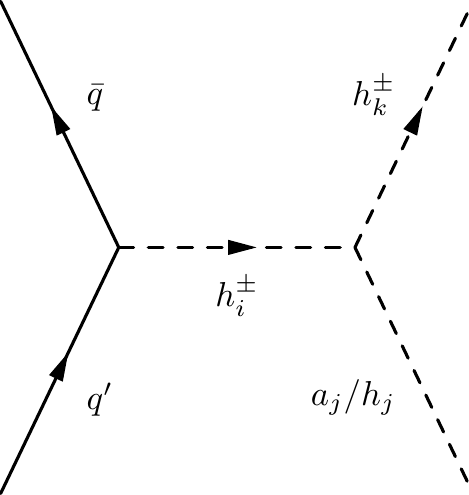}}\hspace{2.3cm}
\subfigure[]{\includegraphics[width=0.15\linewidth]{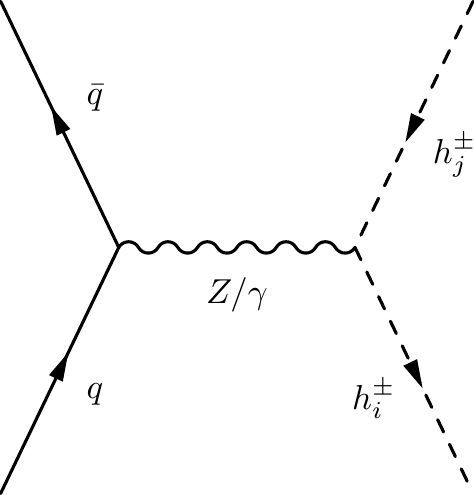}}}
\captionof{figure}{Relevant Feynman diagrams for the production of charged Higgs boson(s).}\label{chprod}
\end{center}
\end{figure}

We have chosen different benchmark points which represent different decay modes preferred either by a triplet-like or by a doublet-like charged Higgs boson \cite{BC3, BC4}. The relevant scalar spectrum for such benchmark points is presented in Table \ref{bench} whereas the dominant production cross-section channels are depicted in Figure \ref{chprod} and their values for each benchmark points are given in Table \ref{cross}.
In particular, BP3 represents a completely doublet-like charged Higgs boson and BP4 represents a  completely triple-like charged Higgs bosons.
In the case of BP1, $a_1 W^\pm$ mode is the most dominant, whereas in BP2 we have mixed scenario.
The triplet-like charged Higgs boson and its decay mode to $ZW^\pm$ can be probed via $3\ell + 1\tau $ with an early data of $\sim 54$ fb$^{-1}$ of integrated luminosity. However the same channel is sensitive to $a_1W^\pm$ with $\sim48$ fb$^{-1}$ of integrated luminosity. In order to distinguish between this two scenarios we have considered the $ \geq 3\tau +\geq 1\ell $ for BP1 ($\sim71$ fb$^{-1}$) and $ \geq4\ell+p^{\ell_1}_T \geq 50\, \rm{GeV}+p^{\ell_2}_T \geq 40\, \rm{GeV} $ for BP4 ($\sim81$ fb$^{-1}$).

\begin{table}
\small
\begin{center}
\begin{tabular}{l l l l l}
\toprule
&$\mathbf{BP1}\qquad\qquad$&$\mathbf{BP2}\qquad\qquad$&$\mathbf{BP3}\qquad\qquad$&$\mathbf{BP4}\qquad\qquad$\\
\midrule
$m_{h_1}\qquad$ & {\color{red}$\sim 125$} & {\color{red}$\sim 125$} & {\color{red}$\sim125$}& {\color{red}$\sim125$}\\
$m_{h_2}$ &$316.14$  & \color{green}$340.44$ &$272.87$& \color{green}$174.21$ \\
$m_{a_1}$ & \color{blue}$41.221$ & \color{blue}$36.145$ &  \color{blue}$30.655$& \color{blue}$61.537$\\
$m_{a_2}$ & \color{green}$181.34$ & $428.68$ & $278.22$ &$1052.7$\\
$m_{h^\pm_1}$ & \color{green}$179.69$ & \color{green}$339.97$ &\color{red}$289.51$& \color{green}$174.11$ \\
\bottomrule
\end{tabular}
\captionof{table}{Scalar spectrum for the benchmark points considered. Masses are expressed in GeV.}\label{bench}
\end{center}
\end{table}

\begin{table}
\small
\begin{center}\label{cross}
\begin{tabular}{l l l l l}
\toprule
&$\mathbf{BP1}\qquad\qquad$&$\mathbf{BP2}\qquad\qquad$&$\mathbf{BP3}\qquad\qquad$&$\mathbf{BP4}\qquad\qquad$\\
\midrule
$h_1^\pm h_1^\mp\qquad$&$148.00$&$13.00$&$12.48$&$166.50$\\
$h_2 h_1^\pm$&$0.28$&$26.42$&$13.89$&$334.62$\\
$a_2 h_1^\pm$&$292.45$&$2.38\times10^{-3}$&$12.54$&$7.07\times10^{-8}$\\
\bottomrule
\end{tabular}
\captionof{table}{Relevant production cross-section, expressed in fb, of $h_1^\pm$ @ the LHC for e.c.m. of 14 TeV.}\label{cross}
\end{center}
\end{table}

\normalsize

\section{331 model}

\begin{figure}[t]
\centering
\mbox{\subfigure[]{
\includegraphics[width=0.225\textwidth]{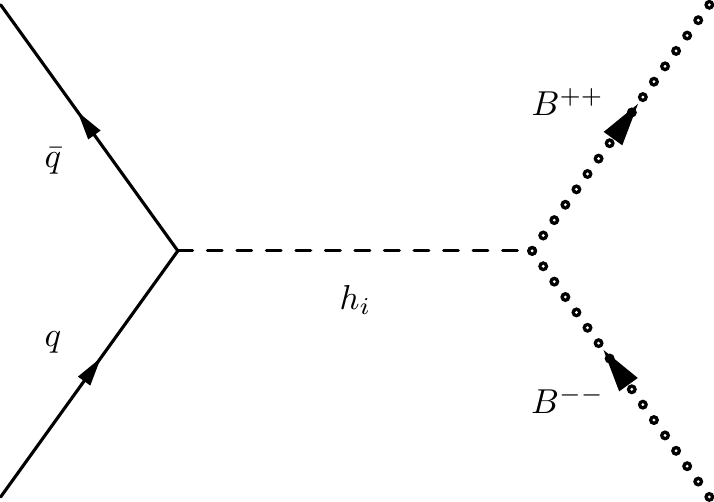}}\hspace{.8cm}
  \subfigure[]{
  \includegraphics[width=0.225\textwidth]{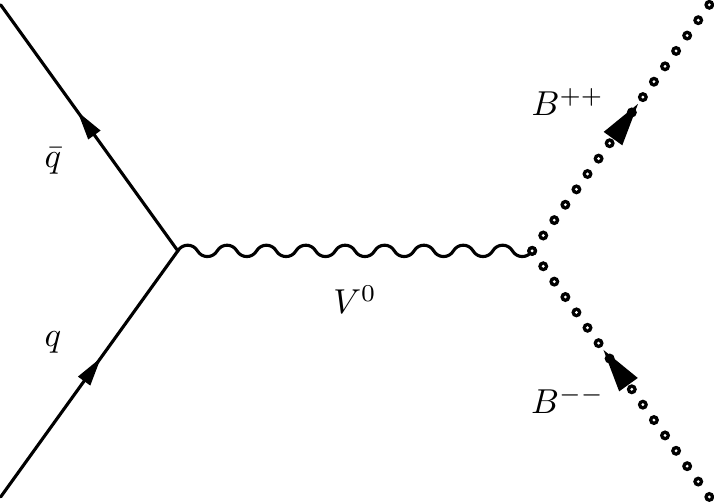}}\hspace{.8cm}
\subfigure[]{\includegraphics[width=0.225\textwidth]{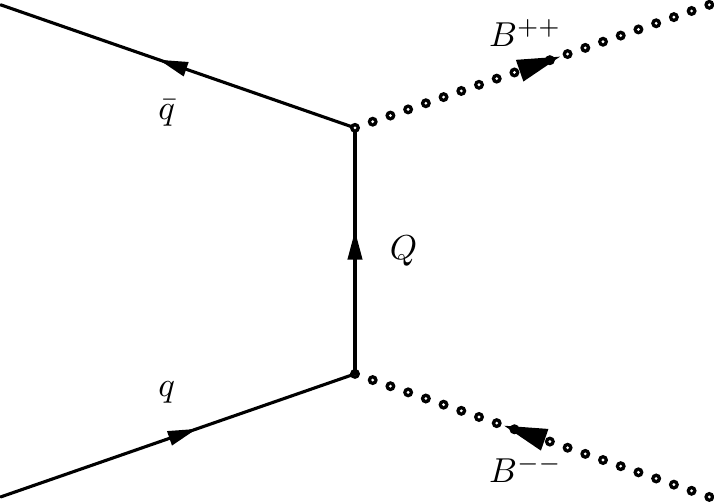}}}
\caption{Typical contributions to events with two doubly-charged bosons in the
  final state and no extra jets.
  (a) and (b): contributions due
  to the mediation of a scalar (a) and a vector (b) boson.
  (c): $t$-channel exchange of exotic quarks $Q$.}
 \label{jetless}
\end{figure}

The minimal version of the 331 model  exhibits
the interesting feature of having both scalar and vector doubly-charged
bosons \cite{PHF, PP, Valle, cccf, cccf2, other331}.
In the context of the minimal 331 model there is an interesting possibility. In fact, one can test whether a same-sign lepton pair has been produced by either a scalar or a vector boson, as shown in Figure \ref{jetless}. The production from scalar boson will also shed light on the presence of a higher representation of the $SU(3)_c\times SU(3)_L\times U(1)_X$ gauge group, namely the sextet.
In the minimal formulation leptons and scalar are triplet of $SU(3)_L$, 

\footnotesize
\begin{equation}
l=\left(
\begin{array}{c}
l_L\\
\nu_l\\
l^c_R
\end{array}
\right)\in(\mathbf{1},\mathbf{\bar 3}, 0),\quad 
\rho=\left(
\begin{array}{c}
\rho^{++}\\
\rho^+\\
\rho^0
\end{array}
\right)\in(\mathbf{1},\mathbf3,1),\quad\eta=\left(
\begin{array}{c}
\eta^+\\
\eta^0\\
\eta^-
\end{array}
\right)\in(\mathbf1,\mathbf3,0),\quad\chi=\left(
\begin{array}{c}
\chi^0\\
\chi^-\\
\chi^{--}
\end{array}
\right)\in(\mathbf1,\mathbf3,-1).
\end{equation}

\normalsize

\noindent
For the complete field content we refer to \cite{cccf, cccf2}. In the 331 the Yukawa for the leptons has to be $\mathcal{L}_{l,\, triplet}^{Yuk}= G^\eta_{a b}\, l^i_{a}\cdot l^j_{b}\,\eta^{* k}\epsilon^{i j k} + \rm{h. c.} $
However the combination $M_{a b}=( l^i_{a}\cdot l^j_{b })\eta^{* k}\epsilon^{i j k}$
is antisymmetric under the exchange of the two flavours, implying
that even $G_{a b}$ has to be antisymmetric, hence is not sufficient to provide mass to all leptons.
We shall solve this problem by introducing a second 
invariant operator, with the inclusion of a sextet $\sigma$:

\small
\begin{equation}
\sigma=\left(
\renewcommand*{\arraystretch}{1.5}
\begin{array}{ccc}
\sigma_1^{++}&\sigma_1^+/\sqrt2&\sigma^0/\sqrt2\\
\sigma_1^+/\sqrt2&\sigma_1^0&\sigma_2^-/\sqrt2\\
\sigma^0/\sqrt2&\sigma_2^-/\sqrt2&\sigma_2^{--}
\end{array}
\right)\in(\mathbf1,\mathbf6,0),
\end{equation}
\normalsize

\noindent
leading to the Yukawa term $\mathcal{L}_{l, sextet}^{{Yuk.}}= G^\sigma_{a b} l^i_a\cdot l^j_b \sigma^*_{i,j}$, with $G^\sigma_{a b}$ symmetric in flavour space. 
If we leave aside the sextet contribution,
the Yukawa for the leptons is related to the scalar triplet $\eta$
which does not possess any doubly-charged state.
This means that revealing a possible decay
$H^{\pm\pm}\to l^\pm l^\pm$ would be a
distinctive signature of the presence of the sextet representation
in the context of the 331 model.

\begin{figure}[t]
\centering
\mbox{\subfigure[]{
\includegraphics[width=0.310\textwidth]{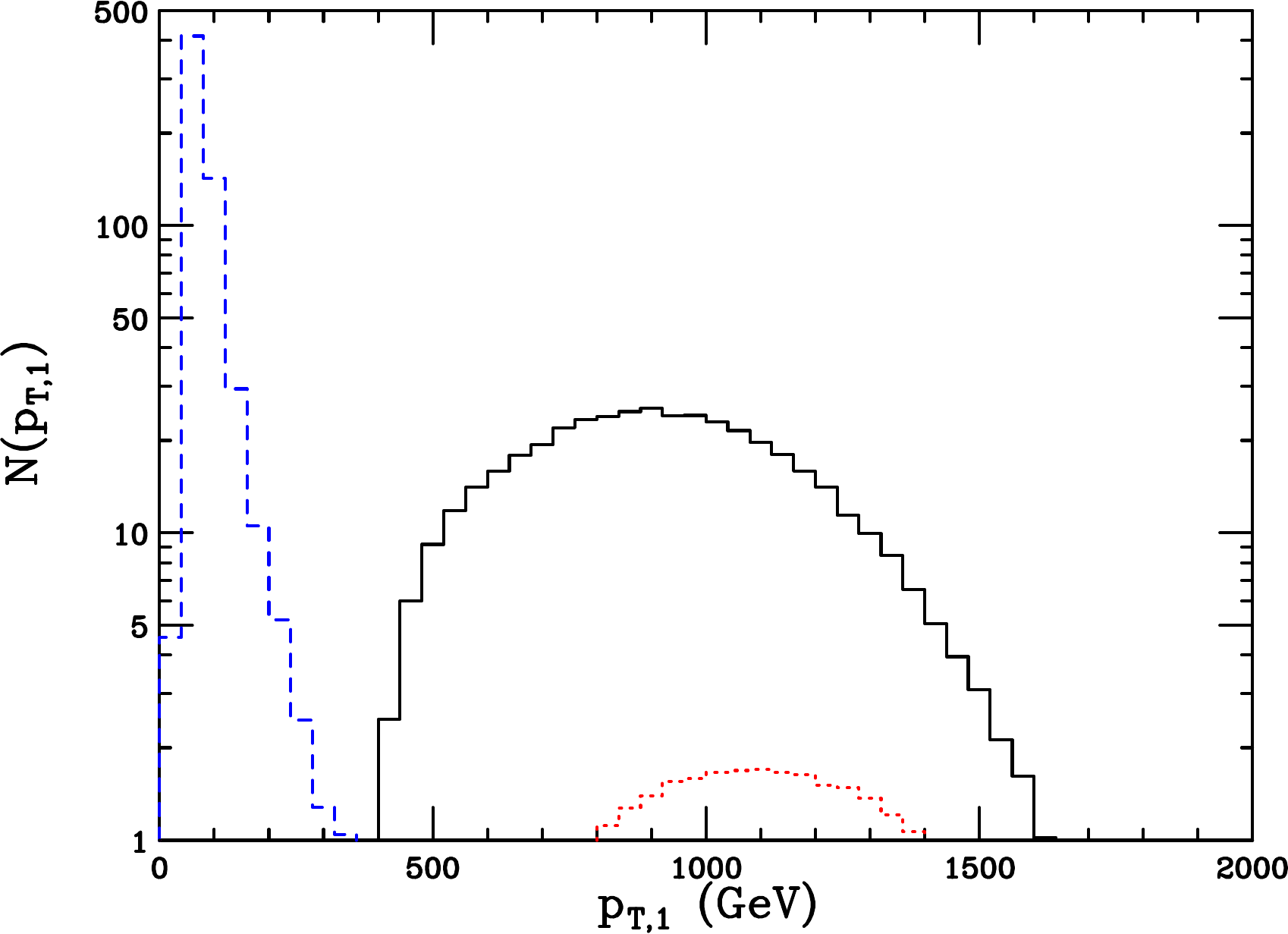}}\hspace{1.5cm}
\subfigure[]{\includegraphics[width=0.310\textwidth]{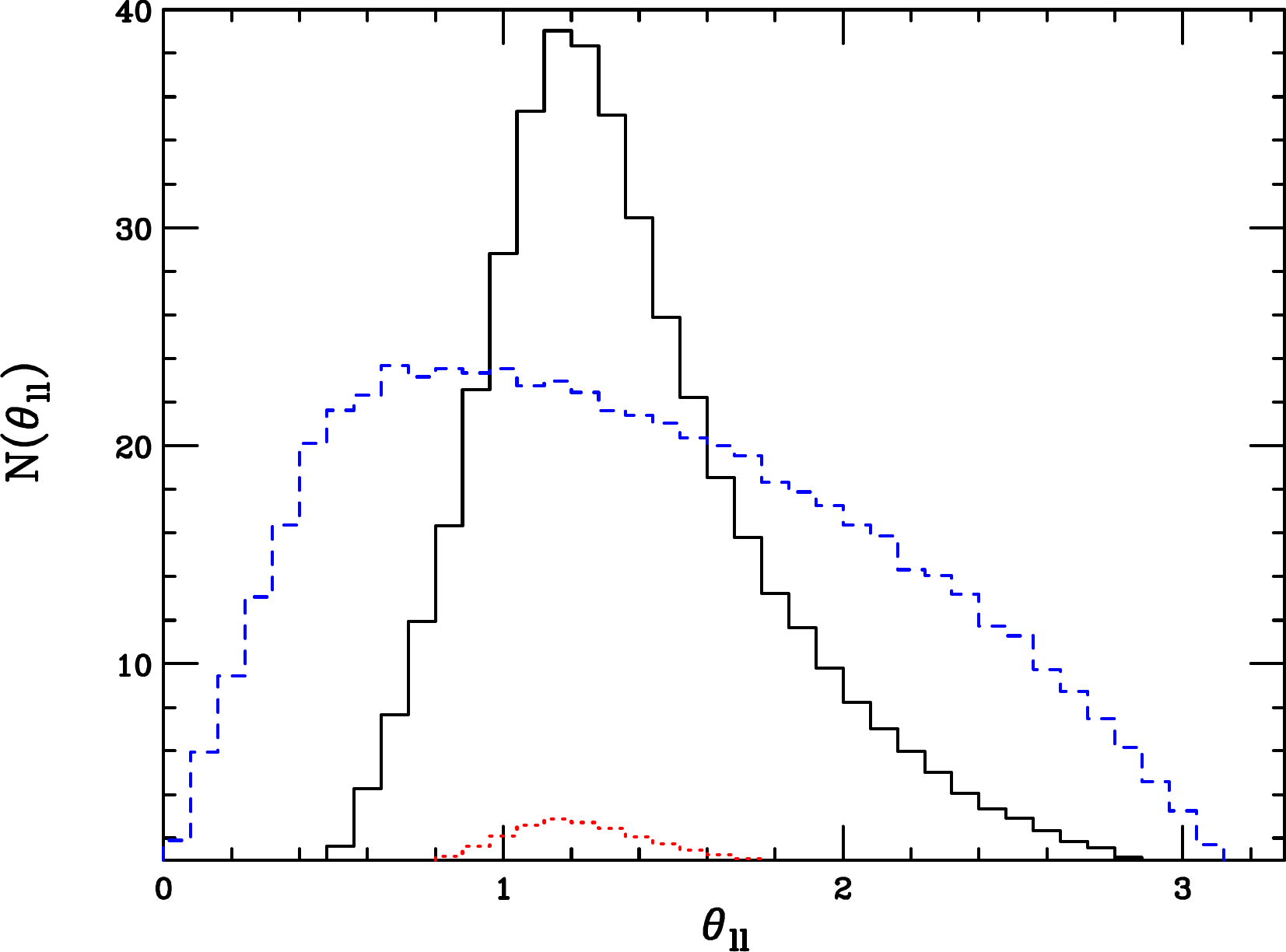}}}
\caption{Distributions of the transverse momentum of the hardest lepton
  (a) and polar angle between same-sign leptons (b).}
 \label{bilep}
\end{figure}

Here we present the main results for the production of
two same-sign lepton pairs at the LHC, mediated by either
vector or scalar bileptons in the 331 model: $pp\to Y^{++}Y^{--}(H^{++}H^{--})\to (l^+l^+)(l^-l^-),
\label{signal}$ where $l=e,\mu$. The distribution of the hardest $p_T$ of the lepton and the polar angle between same-sign leptons are shown in Figure \ref{bilep}.
At 13 TeV LHC, 
after cuts are applied, the LO cross sections, computed by 
\texttt{MadGraph}, read $\sigma(pp\to YY\to 4l)\simeq
4.3~{\rm fb}\ ;\ \sigma(pp\to HH\to 4l)\simeq 0.3~{\rm fb}.$ The difference in the cross sections can be explained in terms of the spin of the intermediate bileptons.

\begin{thebibliography}{99}

\bibitem{ATLASdisc}
  G.~Aad {\it et al.} [ATLAS Collaboration],
  {\it Observation of a new particle in the search for the Standard Model Higgs boson with the ATLAS detector at the LHC}.
  Phys.\ Lett.\ B {\bf 716} (2012) 1.
  
\bibitem{CMSdisc}
  S.~Chatrchyan {\it et al.} [CMS Collaboration],
  {\it Observation of a new boson at a mass of 125 GeV with the CMS experiment at the LHC}.
  Phys.\ Lett.\ B {\bf 716} (2012) 30.
  
\bibitem{BC1}
  P.~Bandyopadhyay, C.~Corian\`o and A.~Costantini,
  {\it Perspectives on a supersymmetric extension of the standard model with a Y = 0 Higgs triplet and a singlet at the LHC}.
  JHEP {\bf 1509} (2015) 045.
  
  
\bibitem{BC2}
  P.~Bandyopadhyay, C.~Corian\`o and A.~Costantini,
  {\it Probing the hidden Higgs bosons of the $Y = 0$ triplet- and singlet-extended Supersymmetric Standard Model at the LHC}.
  JHEP {\bf 1512} (2015) 127.
  
  
\bibitem{BC3}
  P.~Bandyopadhyay, C.~Corian\`o and A.~Costantini,
  {\it General analysis of the charged Higgs sector of the $Y=0$ triplet-singlet extension of the MSSM at the LHC}.
  Phys.\ Rev.\ D {\bf 94} (2016) no.5,  055030.
  
  
\bibitem{BC4}
  P.~Bandyopadhyay and A.~Costantini,
  {\it Distinguishing charged Higgs bosons from different representations at the LHC}.
  JHEP {\bf 1801} (2018) 067.  

\bibitem{PHF}
P.H. Frampton, 
{\it Chiral Dilepton Model and the Flavor Question.}
Phys. Rev. Lett. {\bf 69,} 2889 (1992). 

\bibitem{PP}
F. Pisano and V. Pleitez, 
{\it An SU(3) X U(1) Model of Electroweak Interactions},
Phys. Rev. {\bf D46,} 410 (1992). 

\bibitem{Valle}
M. Singer, J.W.F. Valle and J. Schechter, {\it Canonical Neutral Current Predictions from the
Weak Electromagnetic Group $SU(3) \times U(1)$},
Phys. Rev. {\bf D22,} 738 (1980). 

  
\bibitem{cccf}
  G.~Corcella, C.~Corian\`o, A.~Costantini and P.~H.~Frampton,
  {\it Bilepton Signatures at the LHC}.
  Phys.\ Lett.\ B {\bf 773} (2017) 544.
  
\bibitem{cccf2}
  G.~Corcella, C.~Corian\`o, A.~Costantini and P.~H.~Frampton,
  {\it Exploring Scalar and Vector Bileptons at the LHC in a 331 Model.}
  Phys.\ Lett.\ B {\bf 785} (2018) 73  

\bibitem{other331}
  Q.~H.~Cao, Y.~Liu, K.~P.~Xie, B.~Yan and D.~M.~Zhang,
  {\it Diphoton excess, low energy theorem, and the 331 model}.
  Phys.\ Rev.\ D {\bf 93} (2016) 075030.

\end{thebibliography}
\end{document}